\documentclass[12pt,aps,preprint]{revtex4}

\usepackage{amsmath}
\usepackage{bm}

\begin{document}


\title{Supersymmetry in the $\bm{O(N)}$ Gross-Neveu Models?}

\author{Herbert Morales}
\email{hmorales@fisica.ucr.ac.cr}
\affiliation{%
Escuela de F\'isica\\
Universidad de Costa Rica\\
San Jos\'e, Costa Rica
}%

\date{\today}

\begin{abstract}
We study a special classical current in the $O(3)$ Gross-Neveu model
that becomes supersymmetric when quantum anomalies are included.
Following its definition, we generalize the current for the general case,
the $O(N)$ Gross-Neveu models.
We compute its algebra and discuss the possibility of supersymmetry
can be established for these models.
\end{abstract}

\maketitle


\newcommand{\spart}{\hbox{$\not \! \partial$}}
\newcommand{\contraction}[2]{%
\vbox{
\hbox{\hskip 1 ex \vbox{\hrule \hbox{\vrule height 0 ex depth 0.5 ex
\hskip #2 \vrule height 0 ex depth 0.5 ex}}}
\vspace{-1.3ex}                        
\hbox{$ #1 $}
} }


\section{Introduction}

The $O(N)$ Gross-Neveu model (GNM) is a renormalizable field theory of an 
$N$-component Majorana Fermi field, transforming in the fundamental 
representation of the orthogonal group $O(N)$, with a quartic 
self-interaction.
This model was first introduced by Nambu and Jona-Lasinio in 
four dimensions as a dynamical model of elementary particles in 
which nucleons and mesons are derived from a fundamental spinor field 
\cite{Nambu:1961tp}.
In two dimensions, Gross and Neveu studied the large $N$ limit and 
performed an expansion in powers of $1/N$ to all orders in the coupling 
constant 
\cite{Gross:1974jv}.
Thus, they found that the model displays dynamical symmetry breakdown
that generates, in the resulting theory, a fermion mass and dimensional 
transmutation, i.e the conversion of a dimensionless coupling constant 
into a mass scale parameter.
All these properties follow from the fact that the $O(N)$ GNM is an
asymptotically free theory.
Subsequently, other features of the model have been found.
In a semiclassical analysis, the particle spectrum of GNM reveals a very 
rich structure 
\cite{Dashen:1975xh}.
The integrability of the classical model due to the existence of an 
infinite number of conservation laws \cite{Neveu:1977cr} (which survive 
quantization) has been used to compute the exact $S$-matrix of the theory 
\cite{Zamolodchikov:1977nu}, \cite{Zamolodchikov:1977hh}.

The Lagrangian for the model is
\begin{equation}
{\cal L} = \frac{i}{2} \bar{\psi}_i \spart \psi_i + g (\bar{\psi}_i 
\psi_i)^2,
\label{Lgnm}
\end{equation}
where $\psi_i$ is a Majorana spinor, $g$ is a dimensionless coupling 
constant and repeated indices are summed over.
A fermion mass term is forbidden by the discrete chiral symmetry 
\begin{equation}
\psi_i \to \gamma^5 \psi_i.
\end{equation}
If we use Dirac spinors in place of Majorana ones, the symmetry group
will be $U(N)$.
In either case, the symmetry that suffers spontaneous breakdown is the 
discrete chiral symmetry.

In the Majorana-Weyl representation, the Lagrangian in (\ref{Lgnm}) can
be written in terms of chiral components of the spinors as ($j \neq i$)
\begin{equation}
{\cal L} = i \psi_{i R} \partial_{+} \psi_{i R} + i \psi_{i L}
\partial_{-} \psi_{i L} + 4 g \psi_{i L} \psi_{j L} \psi_{i R} \psi_{j R},
\end{equation}
and so can the field equations
\begin{equation}
\partial_{+} \psi_{i R} = 4 i g \psi_{j L} \psi_{j R} \psi_{i L},
\qquad
\partial_{-} \psi_{i L} = -4 i g \psi_{j L} \psi_{j R} \psi_{i R}.
\label{eom}
\end{equation}

In this paper, we investigate if the $O(N)$ GNM can have
supercurrents at the quantum level.
This work is organized as follows.
In section \ref{sec:gnm3},  we define and study a special
classical current in the $O(3)$ GNM that becomes supersymmetric
when quantum anomalies are included.
With its modified current, we compute the superalgebra.
In section \ref{sec:gnm6}, we generalize the current definition
to the $O(N)$ GNM and compute the new algebra.
Then we conclude with remarks on the supersymmetric
interpretation of this current and its algebra.


\section{\label{sec:gnm3} The $\bm{O(3)}$ Gross-Neveu Model}

Following \cite{Witten:1977xv}, let us define the following  fermionic 
currents which will later become the supercurrents
\begin{equation}
j^{+} \equiv \psi_{1 R} \psi_{2 R} \psi_{3 R},
\qquad
j^{-} \equiv \psi_{1 L} \psi_{2 L} \psi_{3 L}.
\end{equation}

Classically, they are conserved
\begin{equation}
\partial_{+} j^{+} = 0, \qquad \partial_{-} j^{-} = 0.
\label{Claw}
\end{equation}
These results follow from the field equations (\ref{eom}) and
$ \psi_{i R}^2 = \psi_{i L}^2 = 0 $.

Strictly, $j^{+}$ and $j^{-}$ form the lower and upper components of the
spinor current, $ J^\mu = \frac{1}{2} (\bar{\psi}_1 \gamma_\nu \psi_2)
\gamma^\nu \gamma^\mu \gamma^5 \psi_3 $, i.e.
\begin{equation}
J^0 = \begin{pmatrix} -j^{-} \cr j^{+} \cr \end{pmatrix}, \qquad
J^1 = \begin{pmatrix}  j^{-} \cr j^{+} \cr \end{pmatrix},
\end{equation}
so that (\ref{Claw}) is implemented in
$ \partial_{\mu} J^{\mu} = 0 $.

At the quantum level, these conservation laws are modified by anomalies.
From dimensional and symmetry considerations, the modifications must
take the form
\begin{equation}
\partial_{\mu} J^{\mu} + f(g) \partial_{\mu} R^{\mu} = 0,
\label{Qlaw}
\end{equation}
where $ R^\mu = \frac{1}{2} (\bar{\psi}_1 \gamma_\nu \psi_2)
\gamma^\mu \gamma^\nu \gamma^5 \psi_3 + (\bar{\psi}_1 \gamma^5 \psi_2)
\gamma^\mu \psi_3$, i.e. 
\begin{equation}
R^0 = \begin{pmatrix} -r^{+} \cr  r^{-} \cr \end{pmatrix}, \qquad
R^1 = \begin{pmatrix} -r^{+} \cr -r^{-} \cr \end{pmatrix},
\end{equation}
\begin{eqnarray}
r^{+} & = & \psi_{1 R} \psi_{2 R} \psi_{3 L} + \psi_{3 R} 
\psi_{1 R}
\psi_{2 L} + \psi_{2 R} \psi_{3 R} \psi_{1 L},
\nonumber \\
r^{-} & = & \psi_{1 L} \psi_{2 L} \psi_{3 R} + \psi_{3 L} 
\psi_{1 L}
\psi_{2 R} + \psi_{2 L} \psi_{3 L} \psi_{1 R},
\end{eqnarray}
and $f(g)$ is some function of the coupling constant.
(See \cite{Witten:1977xv} for a complete discussion of this result.)

Now we define the modified current of $J^\mu$ which is quantumly
conserved by (\ref{Qlaw})
\begin{equation}
{\cal J}^\mu \equiv E(g) ( J^\mu + f(g) R^\mu ),
\end{equation}
where $E(g)$ is a normalization coefficient.
The spinor charge associated with this current is defined in the
usual way
\begin{equation}
Q \equiv \int dx\, {\cal J}^0 = E(g) \int dx\, (J^0 + f(g) R^0),
\end{equation}
or in components, $Q = ( -Q^{-}\ Q^{+} )^T$,
\begin{equation}
Q^{\pm} = E(g) \int dx\, ( j^{\pm} + f(g) r^{\mp} ).
\end{equation}

Using contractions (see Appendix \ref{ap:contraction}), we compute the 
algebra made by these charges
\begin{eqnarray}
(Q^{+})^2 & = & \frac{E^2(g)}{2 \pi}\, (1 - f^2(g))\! \int\! dx\, \bigg(
\frac{i}{2}\, \psi_{i R} \partial_{1} \psi_{i R} + \frac{\pi f(g)}{1 +
f(g)}\, \psi_{i L} \psi_{j L} \psi_{i R} \psi_{j R} \bigg),
\nonumber \\
(Q^{-})^2 & = & \frac{E^2(g)}{2 \pi}\, (1 - f^2(g))\! \int\! dx\, \bigg(
\frac{-i}{2}\, \psi_{i L} \partial_{1} \psi_{i L} + \frac{\pi f(g)}{1 +
f(g)}\, \psi_{i L} \psi_{j L} \psi_{i R} \psi_{j R} \bigg),
\nonumber \\
\{ Q^{+}, Q^{-} \} & = & 0.
\label{susyO3}
\end{eqnarray}
(We have omitted a c-number in the first two equations.) 
For this algebra to be supersymmetric, we need the integrals to be
proportional to the conserved charges associated with translations,
$P^\pm = P^0 \pm P^1 = \int dx\, (T^{0 0} \pm T^{0 1})$, respectively.
Therefore we read off
\begin{eqnarray}
T^{00} & = & \frac{E^2(g)}{4 \pi}\, (f^2(g) - 1) \bigg(
\frac{i}{2} (\psi_{i L} \partial_1 \psi_{i L} - \psi_{i R} \partial_1
\psi_{i R})
- 2 \pi \left( \frac{f(g)}{1 + f(g)} \right)
\psi_{i L} \psi_{j L} \psi_{i R} \psi_{j R} \bigg),
\nonumber \\
T^{01} & = & \frac{E^2(g)}{4 \pi}\, (f^2(g) - 1) \bigg( \frac{-i}{2}
(\psi_{i L} \partial_1 \psi_{i L} + \psi_{i R} \partial_1 \psi_{i R})
\bigg).
\end{eqnarray}
These components of $T^{\mu \nu}$ are expected to be the quantum ones.
(Notice that the expressions in the big parentheses agree with the
classical components if $ f(g)/(1+f(g)) = 2g/\pi $.)
Now $E(g)$ can be found through the proper normalization of
$T^{\mu \nu}$ and $f(g)$ will encode the trace anomaly,
$T^{\mu}_{\ \mu} \neq 0$.
However, for our purpose we do not need to determine the exact form of 
these functions of $g$ (see next paragraph).
The conservation of this stress-energy tensor follows from supersymmetry.

Moreover, the third equation in (\ref{susyO3}) says that the central
charge for the model is zero (independent of the coupling constant).
The exact form of $f(g)$ is just important for the trace anomaly,
but not for the central charge.
(See \cite{Morales:2005} for a discussion of the equivalence between
this model and the supersymmetric sine-Gordon model.)


\section{\label{sec:gnm6} The $\bm{O(N)}$ Gross-Neveu Model}

As the above section, let us define the following fermionic 
currents for $N>3$
\begin{equation}
j^{+}_{ijk} \equiv \psi_{i R} \psi_{j R} \psi_{k R},
\qquad
j^{-}_{ijk} \equiv \psi_{i L} \psi_{j L} \psi_{k L},
\end{equation}
where normal ordering should be understood on the right-hand side of
these equations.
A more proper definition for these currents can be established by
using the $O(N)$ totally antisymmetric tensor
\begin{eqnarray}
s^{+}_{i_1 i_2 \ldots i_{N-3}} & \equiv &
     \frac{1}{3!}\, \epsilon_{i_1 i_2 \ldots i_N}\,
     \psi_{i_{N-2} R} \psi_{i_{N-1} R} \psi_{i_{N} R} =
     \frac{1}{3!}\, \epsilon_{i_1 i_2 \ldots i_N}\,
     j^{+}_{i_{N-2} i_{N-1} i_N},
\nonumber \\
s^{-}_{i_1 i_2 \ldots i_{N-3}} & \equiv &
     \frac{1}{3!}\, \epsilon_{i_1 i_2 \ldots i_N}\,
     \psi_{i_{N-2} L} \psi_{i_{N-1} L} \psi_{i_{N} L} =
     \frac{1}{3!}\, \epsilon_{i_1 i_2 \ldots i_N}\,
     j^{-}_{i_{N-2} i_{N-1} i_N}.
\end{eqnarray}
Now the interpretation of $j^{+}_{ijk}$ is easier: these
currents form the components of a $O(N)$ antisymmetric pseudotensor of
rank $N-3$ (a pseudovector for $N=4$ and a pseudoscalar for $N=3$)
and there are $N(N-1)(N-2)/6$ independent nonzero tensor components
(or currents).
A similar interpretation follows for the currents $j^{-}_{ijk}$.
However, we would rather use $j^{\pm}_{ijk}$ than
$s^{\pm}_{i_1 i_2 \ldots i_{N-3}}$ because the index notation is
simpler.

Classically, they are not conserved
\begin{eqnarray}
\partial_{+} j^{+}_{i j k} & = & 4 i g\, \psi_{l L} \psi_{l R}\,
r^{+}_{i j k},
\nonumber \\
\partial_{-} j^{-}_{i j k} & = & - 4 i g\, \psi_{l L} \psi_{l R}\,
r^{-}_{i j k},
\label{Cijk}
\end{eqnarray}
where the sum over $l$ does not include the values of $i$, $j$ and $k$
and the ``anomalies" are
\begin{eqnarray}
r^{+}_{ijk} & = & \psi_{i R} \psi_{j R} \psi_{k L} + \psi_{k R} 
\psi_{i R} \psi_{j L} + \psi_{j R} \psi_{k R} \psi_{i L},
\nonumber \\
r^{-}_{ijk} & = & \psi_{i L} \psi_{j L} \psi_{k R} + \psi_{k L} 
\psi_{i L} \psi_{j R} + \psi_{j L} \psi_{k L} \psi_{i R}.
\end{eqnarray}
The formal definition for these anomalies is given by
\begin{eqnarray}
t^{+}_{i_1 i_2 \ldots i_{N-3}} & \equiv &
     \frac{1}{2!}\, \epsilon_{i_1 i_2 \ldots i_N}\,
     \psi_{i_{N-2} R} \psi_{i_{N-1} R} \psi_{i_N L} =
     \frac{1}{3!}\, \epsilon_{i_1 i_2 \ldots i_N}\,
     r^{+}_{i_{N-2} i_{N-1} i_N},
\nonumber \\
t^{-}_{i_1 i_2 \ldots i_{N-3}} & \equiv &
     \frac{1}{2!}\, \epsilon_{i_1 i_2 \ldots i_N}\,
     \psi_{i_{N-2} L} \psi_{i_{N-1} L} \psi_{i_N R} =
     \frac{1}{3!}\, \epsilon_{i_1 i_2 \ldots i_N}\,
     r^{-}_{i_{N-2} i_{N-1} i_N}.
\end{eqnarray}

Strictly, $j^{+}_{i j k}$ and $j^{-}_{i j k}$ form the lower and upper
components of the spinor current,
$J^\mu_{i j k} = \frac{1}{2} (\bar{\psi}_i \gamma_\nu \psi_j)
\gamma^\nu \gamma^\mu \gamma^5 \psi_k $, i.e.
\begin{equation}
J^0_{i j k} = \begin{pmatrix}
       -j^{-}_{i j k} \cr j^{+}_{i j k} \cr \end{pmatrix}, \qquad
J^1_{i j k} = \begin{pmatrix}  
        j^{-}_{i j k} \cr j^{+}_{i j k} \cr \end{pmatrix},
\end{equation}
and so do $r^{+}_{i j k}$ and $r^{-}_{i j k}$ for the spinor anomaly,
$R^\mu_{i j k} = \frac{1}{2} (\bar{\psi}_i \gamma_\nu \psi_j)
\gamma^\mu \gamma^\nu \gamma^5 \psi_k + (\bar{\psi}_i \gamma^5 \psi_j)
\gamma^\mu \psi_k$, i.e. 
\begin{equation}
R^0_{i j k} = \begin{pmatrix}
       -r^{+}_{i j k} \cr  r^{-}_{i j k} \cr \end{pmatrix}, \qquad
R^1_{i j k} = \begin{pmatrix}
       -r^{+}_{i j k} \cr -r^{-}_{i j k} \cr \end{pmatrix}.	
\end{equation}
so that (\ref{Cijk}) is implemented in $ \partial_{\mu} J^{\mu}_{ijk} +
2 i g\, ( \bar{\psi}_l \psi_l ) \gamma_{\mu} R^\mu_{ijk} = 0 $.

At the quantum level, we redefine these currents with their anomalies as
\begin{equation}
{\cal J}^\mu_{i j k} \equiv E_N ( J^\mu_{i j k} + f_N\, R^\mu_{i j k} ),
\end{equation}
where $f_N = f_N (g)$ is some function of the coupling constant and
$E_N = E_N (g)$ is a normalization coefficient.
Obviously, different $N$ implies different $f_N$ and $E_N$, but these
functions do not change for fixed $N$ and $g$, due to their $O(N)$
invariance.

The spinor charges associated with these currents are defined by
\begin{equation}
Q_{i j k} \equiv \int dx\, {\cal J}^0_{i j k} =
     E_N\! \int dx\, (J^0_{i j k} + f_N\, R^0_{i j k}),
\end{equation}
or in components, $Q_{i j k} = ( -Q^{-}_{i j k}\ Q^{+}_{i j k} )^T$,
\begin{equation}
Q^{\pm}_{i j k} = E_N\! \int dx\, ( j^{\pm}_{i j k} +
     f_N\, r^{\mp}_{i j k} ).
\end{equation}
However, we can define formally these spinor charges with
\begin{equation}
{\cal Q}^{\pm}_{i_1 i_2 \ldots i_{N-3}} \equiv
    E_N\! \int dx\, ( s^{\pm}_{i_1 i_2 \ldots i_{N-3}} +
            f_N\, t^{\mp}_{i_1 i_2 \ldots i_{N-3}} )
  = \frac{1}{3!}\, \epsilon_{i_1 i_2 \ldots i_N}\,
     Q^{\pm}_{i_{N-2} i_{N-1} i_N}.
\end{equation}

Using the $Q^{\pm}_{i j k}$ algebra (see Appendix \ref{ap:contraction}), we
compute the algebra made by the contracted
${\cal Q}^{\pm}_{i_1  i_2 \ldots i_{N-3}}$
\begin{eqnarray}
{\cal Q}^{+}_{i_1 i_2 \ldots i_{N-3}} {\cal Q}^{+}_{i_1 i_2 \ldots i_{N-3}}
   & = & \frac{(N-1)!}{4 \pi}\, E^2_N\, (1 - f^2_N)\! \int\! dx\,
\bigg( \frac{i}{2}\, \psi_{i R} \partial_{1} \psi_{i R} +
\nonumber \\
   &   & \frac{2 \pi}{N - 1} \left( \frac{f_N}{1 + f_N} \right)
     \psi_{i L} \psi_{j L} \psi_{i R} \psi_{j R} \bigg),
\nonumber \\
{\cal Q}^{-}_{i_1 i_2 \ldots i_{N-3}} {\cal Q}^{-}_{i_1 i_2 \ldots i_{N-3}}
   & = & \frac{(N-1)!}{4 \pi}\, E^2_N\, (1 - f^2_N)\! \int\! dx\,
\bigg( \frac{-i}{2} \psi_{i L} \partial_{1} \psi_{i L} +
\nonumber \\
   &   & \frac{2 \pi}{N - 1} \left( \frac{f_N}{1 + f_N} \right)
     \psi_{i L} \psi_{j L} \psi_{i R} \psi_{j R} \bigg),
\nonumber \\
\{ {\cal Q}^{+}_{i_1 i_2 \ldots i_{N-3}},\,
   {\cal Q}^{-}_{i_1 i_2 \ldots i_{N-3}} \} & = & 0.
\label{susyO6}
\end{eqnarray}
(We have again omitted a c-number in the first two equations.) 
For this algebra to be ``supersymmetric'', we require the integrals to be
proportional to the conserved charges, $P^\pm$, respectively.
Therefore we read off
\begin{eqnarray}
T^{00} & = & \frac{(N-1)!}{8 \pi}\, E_N^2\, (f_N^2 - 1) \bigg( \frac{i}{2}
(\psi_{i L} \partial_1 \psi_{i L} - \psi_{i R} \partial_1 \psi_{i R})
- \frac{4 \pi}{N - 1} \left( \frac{f_N}{1 + f_N} \right)
\psi_{i L} \psi_{j L} \psi_{i R} \psi_{j R} \bigg),
\nonumber \\
T^{01} & = & \frac{(N-1)!}{8 \pi}\, E_N^2\, (f_N^2 - 1) \bigg( \frac{-i}{2}
(\psi_{i L} \partial_1 \psi_{i L} + \psi_{i R} \partial_1 \psi_{i R})
\bigg).
\end{eqnarray}
We try a similar interpretation of these results as those in the above section:
these components of $T^{\mu \nu}$ are expected to be the quantum ones.
(Notice again that the classical expressions can be obtained if
$ f_N /(1 + f_N)(N-1) = g/\pi $.)
Consequently, $E_N$ can be found through the proper normalization of
$T^{\mu \nu}$ and $f_N$ will encode the trace anomaly.
If we consider the third equation in (\ref{susyO6}) as usual, we conclude that
the central charges for the $O(N)$ Gross-Neveu models are zero and
independent of the coupling constant $g$,
the exact forms of $E_N$ and $f_N$ are not important for finding the
``central charges".
The conservation of this stress-energy tensor cannot follow from
``supersymmetry", since we have not shown the current conservation
for ${\cal J}^\mu_{i j k}$.
We expect that the current conservation occurs only for some values
of $g$, because dimensional and symmetry considerations do not
allow us to constraint or infer it.
Thus, we shall have to explore the currents with more details in the
future.

As has been seen, our proposal for ``supersymmetry" requires to
contract the charges ${\cal Q}^{\pm}_{i_1  i_2 \ldots i_{N-3}}$,
otherwise their algebra is very complicated
(see Appendix \ref{ap:contraction}).
This requirement does not allow us to make a direct identification 
with (extended) supersymmetry.
Further research of this algebra is needed to understand its
implications and supersymmetry connections.
Obviously, one can start it with the free models ($g=0$), but
even though the expressions are simpler, the contraction
requirement is still needed.


\appendix

\section{Notation}

We use the metric $ \eta^{0 0} = - \eta^{1 1}  = 1 $
and the antisymmetric tensor
$ \epsilon_{0 1} = - \epsilon^{0 1} = 1 $.
In light-cone coordinates, we define
$ x^{\pm} \equiv x^0 \pm x^1 $,
thus
$ \eta_{+-} = 1/2 $, $\eta^{+-} = 2 $ 
and
$ \epsilon_{+-} = -1/2 $, $ \epsilon^{+-} = 2 $.
Moreover,
$ \partial_{\pm} \equiv (\partial_0 \pm \partial_1)/2 $.

For the two-dimensional Dirac algebra,
we define
$ \gamma^5 = \gamma_5 \equiv \frac{1}{2} \epsilon_{\mu \nu} \gamma^\mu
\gamma^\nu $.
Some useful properties of the $\gamma$ matrices are
$ \gamma^{\mu} \gamma^5 = \epsilon^{\mu \nu} \gamma_{\nu} $ and
$ \gamma^{\mu} \gamma^{\nu} = \eta^{\mu \nu} I - \epsilon^{\mu \nu}
\gamma^5 $.
In the Majorana-Weyl representation, the $\gamma$ matrices are written as
$ \gamma^0 = \sigma_2 $,
$ \gamma^1 = -i \sigma_1 $ and
$ \gamma^5 = -\sigma_3 $.
In light-cone coordinates, we have
$ \gamma^{\pm} \equiv \gamma^0 \pm \gamma^1 $.
Furthermore, a Majorana spinor is written in chiral components as
$ \psi = ( \psi_L \ \psi_R )^T $,
with $\psi_L^{\dag} = \psi_L$ and $\psi_R^{\dag} = \psi_R$.
And a Dirac spinor is given by two Majorana spinors,
$ \psi = (\psi_1 + i \psi_2)/\sqrt{2} $.

Some useful delta-function properties are
$\delta^{(n)} (x - y) = (-1)^{n} \delta^{(n)} (y - x)$ and
\begin{equation}
\int\! dx\, f(x) \delta^{(n)} (x - y) = (-1)^{n} \partial^{n} f(y),
\end{equation}
all derivatives are respect to the argument.


\section{\label{ap:contraction} Fermi fields}

The contraction of free massless Dirac fields is given by the Wightman 
function (see \cite{Klaiber:1968}, \cite{Abdalla:1991vu})
\begin{equation}
S^{(+)} (\xi = y - x) \equiv \contraction{\psi (y) \bar{\psi} (x)}{2 em}
   = \frac{1}{2 \pi} \int \frac{dk^1}{2 k^0} \, k^\mu \gamma_\mu \, 
e^{-i k \cdot \xi - k^0 \epsilon},
\end{equation}
where $\epsilon > 0$ is a UV regularization.

We define
\begin{equation}
C^{\mu} (\xi) \equiv \frac{1}{2} {\rm tr} (\gamma^{\mu} S^{(+)} (\xi))
   = \frac{1}{4 \pi} \int dk^1\, {\rm sgn} (k^\mu) e^{-i k \cdot \xi -
k^0 \epsilon},
\end{equation}
so that $S^{(+)} (\xi) = C^{\mu} (\xi) \gamma_{\mu}$.

In the Majorana-Weyl representation, we have
\begin{eqnarray}
C^{+} (\xi = y - x) & = & \contraction{\psi_R (y) \psi_R (x)}{2.5 em} =
- \frac{i}{2 \pi} \bigg( \frac{1}{\xi^{-} - i \epsilon} \bigg),
\nonumber \\
C^{-} (\xi = y - x) & = & \contraction{\psi_L (y) \psi_L (x)}{2.5 em} =
- \frac{i}{2 \pi} \bigg( \frac{1}{\xi^{+} - i \epsilon} \bigg),
\\
\contraction{\psi_R (y) \psi_L (x)}{2.5 em} & = &
\contraction{\psi_L (y) \psi_R (x)}{2.5 em} = 0,
\nonumber
\end{eqnarray}
where $\psi_R$ and $\psi_L$ are the chiral components of a Majorana 
spinor.

For equal-time situations (from here onward $x$ denotes just the space 
coordinate), we have the following properties
\begin{eqnarray}
C^{\pm} (-x) & = & C^{\mp} (x),
\nonumber \\
\Big( C^{\pm} (x) \Big)^n & = & \bigg( \mp \frac{i}{2 \pi} \bigg)^{n-1}
\frac{1}{(n - 1)!} \Big( \partial_{x}^{(n - 1)} C^{\pm} (x) \Big),
\nonumber \\
\Big( C^{\pm} (x) \Big)^n + (-1)^{n-1} \Big( C^{\pm} (-x) \Big)^n & = &
\bigg( \mp \frac{i}{2 \pi} \bigg)^{n - 1} \frac{1}{(n - 1)!}
\Big( \partial_{x}^{(n - 1)} \delta (x) \Big),
\nonumber \\
C^{-} (x) C^{+} (x) - C^{-} (-x) C^{+} (-x) & = & 0,
\\
(C^{\pm} (x))^2 C^{\mp} (x) + (C^{\pm} (-x))^2 C^{\mp} (-x) & = &
- \frac{1}{8 \pi^2} \frac{\delta' (x)}{x}. 
\nonumber
\end{eqnarray}

Using these functions $C^{\pm}$ and Wick's theorem, we can
``derive" the equal-time canonical anticommutation relations for the
Majorana-Weyl spinor components
\begin{eqnarray}
\{ \psi_R (x), \psi_R (y) \} & = & C^{+} (x - y) + C^{+} (y - x) = \delta 
(x - y),
\nonumber \\
\{ \psi_L (x), \psi_L (y) \} & = & C^{-} (x - y) + C^{-} (y - x) = \delta 
(x - y),
\\
\{ \psi_R (x), \psi_L (y) \} & = & 0.
\nonumber
\end{eqnarray}

Similarly, we can compute other equal-time commutators and anticommutators
of these components.
For example, we find
\begin{eqnarray}
\Big[ (\psi_{1 R} \psi_{2 R}) (x), (\psi_{1 R} \psi_{2 R}) (y) \Big]
    & = & - (C^{+} (x-y))^2 + (C^{+} (y-x))^2
\nonumber \\
    & = & \frac{i}{2 \pi} \delta' (x-y),
\nonumber \\
\Big[ (\psi_{1 L} \psi_{2 L}) (x), (\psi_{1 L} \psi_{2 L}) (y) \Big]
    & = & - \frac{i}{2 \pi} \delta' (x-y),
\\
\Big[ (\psi_{1 R} \psi_{2 R}) (x), (\psi_{1 L} \psi_{2 L}) (y) \Big]
    & = & 0.
\nonumber
\end{eqnarray}
With the above results and
\begin{equation}
j^{0} = i ( \psi_{1 R} \psi_{2 R} + \psi_{1 L} \psi_{2 L} ), \qquad
j^{1} = i ( \psi_{1 R} \psi_{2 R} - \psi_{1 L} \psi_{2 L} ),
\end{equation}
we can obtain the nonzero current commutator
\begin{equation}
\, [ j^{0} (x), j^{1} (y) ] = - \frac{i}{\pi} \delta' (x-y),
\end{equation}
where the right-hand side is a Schwinger term 
\cite{Schwinger:1959xd}.
One can also show the Kac-Moody algebra made by the fermion currents
$j^+_{i j} = 2 i  \psi_{i R} \psi_{j R}$ and
$j^-_{i j} = 2 i  \psi_{i L} \psi_{j L}$
\cite{Witten:1983ar}.

For finding the anticommutators of $Q^{\pm}_{ijk}$, we need the
equal-time anticommutators of three-spinor products.
The calculations are done in the same fashion as before, so that
normal ordering should be understood on the right-hand side of the
following equations.
There is no sum over any index except for $l$ and $m$ 
which run only for the values of $i,j,k$ and all indices are different
from each other.

Hence, the nontrivial three-spinor anticommutators  (those that involve
more than one contraction) needed for
$Q^{+}_{ijk}$ are
\begin{eqnarray}
\{ (\psi_{i R} \psi_{j R} \psi_{k R}) (x), 
   (\psi_{i R} \psi_{j R} \psi_{k R}) (y) \} & = & 
\frac{i}{2 \pi} \delta' (x - y) \psi_{l R} (x) \psi_{l R} (y)
+ \frac{1}{8 \pi^2} \delta'' (x - y),
\nonumber \\
\{ (\psi_{i R} \psi_{j R} \psi_{k R}) (x),
   (\psi_{i R} \psi_{j R} \psi_{k' R}) (y) \} & = &
\frac{i}{2 \pi} \delta' (x - y) \psi_{k R} (x) \psi_{k' R} (y),
\nonumber \\
\{ (\psi_{i R} \psi_{j L} \psi_{k L}) (x),
   (\psi_{i R} \psi_{j L} \psi_{k L}) (y) \} & = &
- \frac{i}{2 \pi} \delta' (x - y) \psi_{i R} (x) \psi_{i R} (y)
+ \frac{1}{8 \pi^2} \frac{\delta' (x - y)}{x - y},
\nonumber \\
\{ (\psi_{i R} \psi_{j L} \psi_{k L}) (x),
   (\psi_{i R} \psi_{j L} \psi_{k' L}) (y) \} & = & 0,
\nonumber \\
\{ (\psi_{i R}  \psi_{j L} \psi_{k L}) (x),
   (\psi_{i' R} \psi_{j L} \psi_{k L}) (y) \} & = &
- \frac{i}{2 \pi} \delta' (x - y) \psi_{i R} (x) \psi_{i' R} (y).
\end{eqnarray}

From these results, we can easily obtain the anticommutators
between $j^{+}_{ijk}$'s
\begin{eqnarray}
\{ j^{+}_{i j k} (x), j^{+}_{i j k} (y) \} & = & 
\frac{i}{2 \pi} \delta' (x - y) \psi_{l R} (x) \psi_{l R} (y)
+ \frac{1}{8 \pi^2} \delta'' (x - y),
\nonumber \\
\{ j^{+}_{i j k} (x), j^{+}_{i j k'} (y) \} & = &
\frac{i}{2 \pi} \delta' (x - y) \psi_{k R} (x) \psi_{k' R} (y),
\nonumber \\
\{ j^{+}_{i j k} (x), j^{+}_{i j' k'} (y) \} & = &
\delta (x - y) (\psi_{j R} \psi_{k R}) (x) (\psi_{j' R} \psi_{k' R}) (y),
\nonumber \\
\{ j^{+}_{i j k} (x), j^{+}_{i' j' k'} (y) \} & = & 0,
\end{eqnarray}
between $j^{+}_{ijk}$ and $r^{-}_{ijk}$
\begin{eqnarray}
\{ j^{+}_{i j k} (x), r^{-}_{i j k} (y) \} & = & \frac{1}{2} \delta (x - y)
  (\psi_{l L} \psi_{m L}) (y) (\psi_{l R} \psi_{m R}) (x),
\nonumber \\
\{ j^{+}_{i j k} (x), r^{-}_{i j k'} (y) \} & = & \delta (x - y) \Big(
  (\psi_{j R} \psi_{k R}) (x) (\psi_{j L} \psi_{k' L}) (y) +
  (\psi_{i R} \psi_{k R}) (x) (\psi_{i L} \psi_{k' L}) (y) \Big),
\nonumber \\
\{ j^{+}_{i j k} (x), r^{-}_{i j' k'} (y) \} & = & \delta (x - y)
  (\psi_{j R} \psi_{k R}) (x) (\psi_{j' L} \psi_{k' L}) (y),
\nonumber \\
\{ j^{+}_{i j k} (x), r^{-}_{i' j' k'} (y) \} & = & 0,
\end{eqnarray}
and between $r^{-}_{ijk}$'s
\begin{eqnarray}
\{ r^{-}_{i j k} (x), r^{-}_{i j k} (y) \} & = & 
- \frac{i}{2 \pi} \delta' (x - y) \psi_{l R} (x) \psi_{l R} (y)
- \delta (x - y) (\psi_{l L} \psi_{m R}) (x) (\psi_{m L} \psi_{l R}) (y) +
\nonumber \\
   &  & + \frac{3}{8 \pi^2} \frac{\delta' (x - y)}{x - y}\, ,
\nonumber \\
\{ r^{-}_{i j k} (x), r^{-}_{i j k'} (y) \} & = & 
- \frac{i}{2 \pi} \delta' (x - y) \psi_{k R} (x) \psi_{k' R} (y)
- \delta (x - y) \Big(
  (\psi_{i R} \psi_{k L}) (x) (\psi_{k' R} \psi_{i L}) (y) +
\nonumber \\
  & & + (\psi_{j R} \psi_{k L}) (x) (\psi_{k' R} \psi_{j L}) (y)
      + (\psi_{k R} \psi_{i L}) (x) (\psi_{i R} \psi_{k' L}) (y) +
\nonumber \\
  & & + (\psi_{k R} \psi_{j L}) (x) (\psi_{j R} \psi_{k' L}) (y) \Big),
\nonumber \\
\{ r^{-}_{i j k} (x), r^{-}_{i j' k'} (y) \} & = & \delta (x - y) \Big(
  (\psi_{j L} \psi_{k L}) (x) (\psi_{j' L} \psi_{k' L}) (y) +
\nonumber \\
   & & + (\psi_{j R} \psi_{k L} - \psi_{k R} \psi_{j L})(x)
      (\psi_{j' R} \psi_{k' L} - \psi_{k' R} \psi_{j' L})(y) \Big),
\nonumber \\
\{ r^{-}_{i j k} (x), r^{-}_{i' j' k'} (y) \} & = & 0.
\end{eqnarray}

Consequently, the algebra for $Q^{+}_{ijk}$ is given by
\begin{eqnarray}
\{ Q^{+}_{i j k}\, , Q^{+}_{i j k} \} & = &
    E^2_N (1 - f_N)\! \int\! dx\, \bigg( 
    \frac{i}{2 \pi} (1 + f_N)\, \psi_{l R} \partial_{1} \psi_{l R}
    + f_N\, \psi_{l L} \psi_{m L} \psi_{l R} \psi_{m R} \bigg)
    + \textrm{c-\#},
\nonumber \\
\{ Q^{+}_{i j k}\, , Q^{+}_{i j k'} \} & = &
    E^2_N (1 - f_N)\! \int\! dx\, \bigg( 
    \frac{i}{4 \pi} (1 + f_N)\, ( \psi_{k R} \partial_{1} \psi_{k' R}
    + \psi_{k' R} \partial_{1} \psi_{k R} ) +
\nonumber \\
   &  & -f_N\, ( \psi_{i L} \psi_{i R} + \psi_{j L} \psi_{j R} )
               ( \psi_{k L} \psi_{k' R} + \psi_{k' L} \psi_{k R} ) \bigg),
\nonumber \\
\{ Q^{+}_{i j k}\, , Q^{+}_{i j' k'} \} & = &
    E^2_N\! \int\! dx\, \bigg( 
    \psi_{j R} \psi_{k R} \psi_{j' R} \psi_{k' R} +
    f_N\, ( \psi_{j L} \psi_{k L} \psi_{j' R} \psi_{k' R} +
            \psi_{j' L} \psi_{k' L} \psi_{j R} \psi_{k R} ) +
\nonumber \\
   &  & f_N^2\, ( \psi_{j L} \psi_{k L} \psi_{j' L} \psi_{k' L} +
       ( \psi_{j L} \psi_{k R} - \psi_{k L} \psi_{j R} )
       ( \psi_{j' L} \psi_{k' R} - \psi_{k' L} \psi_{j' R} ) ) \bigg),
\nonumber \\
\{ Q^{+}_{i j k}\, , Q^{+}_{i' j' k'} \} & = & 0.
\end{eqnarray}

In similar fashion, we construct the $Q^{-}_{ijk}$ algebra
\begin{eqnarray}
\{ Q^{-}_{i j k}\, , Q^{-}_{i j k} \} & = &
    E^2_N (1 - f_N)\! \int\! dx\, \bigg( 
    \frac{- i}{2 \pi} (1 + f_N)\, \psi_{l L} \partial_{1} \psi_{l L}
    + f_N\, \psi_{l L} \psi_{m L} \psi_{l R} \psi_{m R} \bigg)
    + \textrm{c-\#},
\nonumber \\
\{ Q^{-}_{i j k}\, , Q^{-}_{i j k'} \} & = &
    E^2_N (1 - f_N)\! \int\! dx\, \bigg(
    \frac{-i}{4 \pi} (1 + f_N)\, ( \psi_{k L} \partial_{1} \psi_{k' L}
    + \psi_{k' L} \partial_{1} \psi_{k L} ) +
\nonumber \\
   &  & -f_N\, ( \psi_{i L} \psi_{i R} + \psi_{j L} \psi_{j R} )
               ( \psi_{k L} \psi_{k' R} + \psi_{k' L} \psi_{k R} ) \bigg),
\nonumber \\
\{ Q^{-}_{i j k}\, , Q^{-}_{i j' k'} \} & = &
    E^2_N\! \int\! dx\, \bigg(
    \psi_{j L} \psi_{k L} \psi_{j' L} \psi_{k' L} +
    f_N\, ( \psi_{j L} \psi_{k L} \psi_{j' R} \psi_{k' R} +
           \psi_{j' L} \psi_{k' L} \psi_{j R} \psi_{k R} ) +
\nonumber \\
   &  & f_N^2\, ( \psi_{j R} \psi_{k R} \psi_{j' R} \psi_{k' R} +
       ( \psi_{j L} \psi_{k R} - \psi_{k L} \psi_{j R} )
       ( \psi_{j' L} \psi_{k' R} - \psi_{k' L} \psi_{j' R} ) ) \bigg),
\nonumber \\
\{ Q^{-}_{i j k}\, , Q^{-}_{i' j' k'} \} & = & 0.
\end{eqnarray}

Finally, we close the $Q^{\pm}_{ijk}$ algebra with
\begin{eqnarray}
\{ Q^{+}_{i j k}\, , Q^{-}_{i j k} \} & = & 0,
\nonumber \\
\{ Q^{+}_{i j k}\, , Q^{-}_{i j k'} \} & = &
    E^2_N\, f_N (1 - f_N)\! \int\! dx\, \bigg(
    ( \psi_{i L} \psi_{i R} + \psi_{j L} \psi_{j R} )
    ( \psi_{k L} \psi_{k' L} + \psi_{k R} \psi_{k' R} ) \bigg),
\nonumber \\
\{ Q^{+}_{i j k}\, , Q^{-}_{i j' k'} \} & = &
    E^2_N\, f_N\! \int\! dx\, \bigg(
    ( \psi_{j R} \psi_{k R} + f_N\, \psi_{j L} \psi_{k L} )
    ( \psi_{j' L} \psi_{k' R} + \psi_{j' R} \psi_{k' L} ) +
\nonumber \\
   &  & ( \psi_{j L} \psi_{k R} + \psi_{j R} \psi_{k L} )
      ( \psi_{j' L} \psi_{k' L} + f_N\, \psi_{j' R} \psi_{k' R} ) \bigg),
\nonumber \\
\{ Q^{+}_{i j k}\, , Q^{-}_{i' j' k'} \} & = & 0.
\end{eqnarray}

We also assume that the $Q^{\pm}_{ijk}$ algebra is still valid in
Gross-Neveu models as a result of their asymptotic freedom.

\bibliography{biblio}

\begin{thebibliography}{12}
\expandafter\ifx\csname natexlab\endcsname\relax\def\natexlab#1{#1}\fi
\expandafter\ifx\csname bibnamefont\endcsname\relax
  \def\bibnamefont#1{#1}\fi
\expandafter\ifx\csname bibfnamefont\endcsname\relax
  \def\bibfnamefont#1{#1}\fi
\expandafter\ifx\csname citenamefont\endcsname\relax
  \def\citenamefont#1{#1}\fi
\expandafter\ifx\csname url\endcsname\relax
  \def\url#1{\texttt{#1}}\fi
\expandafter\ifx\csname urlprefix\endcsname\relax\def\urlprefix{URL }\fi
\providecommand{\bibinfo}[2]{#2}
\providecommand{\eprint}[2][]{\url{#2}}

\bibitem[{\citenamefont{Nambu and Jona-Lasinio}(1961)}]{Nambu:1961tp}
\bibinfo{author}{\bibfnamefont{Y.}~\bibnamefont{Nambu}} \bibnamefont{and}
  \bibinfo{author}{\bibfnamefont{G.}~\bibnamefont{Jona-Lasinio}},
  \bibinfo{journal}{Phys. Rev.} \textbf{\bibinfo{volume}{122}},
  \bibinfo{pages}{345} (\bibinfo{year}{1961}).

\bibitem[{\citenamefont{Gross and Neveu}(1974)}]{Gross:1974jv}
\bibinfo{author}{\bibfnamefont{D.~J.} \bibnamefont{Gross}} \bibnamefont{and}
  \bibinfo{author}{\bibfnamefont{A.}~\bibnamefont{Neveu}},
  \bibinfo{journal}{Phys. Rev.} \textbf{\bibinfo{volume}{D10}},
  \bibinfo{pages}{3235} (\bibinfo{year}{1974}).

\bibitem[{\citenamefont{Dashen et~al.}(1975)\citenamefont{Dashen, Hasslacher,
  and Neveu}}]{Dashen:1975xh}
\bibinfo{author}{\bibfnamefont{R.~F.} \bibnamefont{Dashen}},
  \bibinfo{author}{\bibfnamefont{B.}~\bibnamefont{Hasslacher}},
  \bibnamefont{and} \bibinfo{author}{\bibfnamefont{A.}~\bibnamefont{Neveu}},
  \bibinfo{journal}{Phys. Rev.} \textbf{\bibinfo{volume}{D12}},
  \bibinfo{pages}{2443} (\bibinfo{year}{1975}).

\bibitem[{\citenamefont{Neveu and Papanicolaou}(1978)}]{Neveu:1977cr}
\bibinfo{author}{\bibfnamefont{A.}~\bibnamefont{Neveu}} \bibnamefont{and}
  \bibinfo{author}{\bibfnamefont{N.}~\bibnamefont{Papanicolaou}},
  \bibinfo{journal}{Commun. Math. Phys.} \textbf{\bibinfo{volume}{58}},
  \bibinfo{pages}{31} (\bibinfo{year}{1978}).

\bibitem[{\citenamefont{Zamolodchikov and
  Zamolodchikov}(1978{\natexlab{a}})}]{Zamolodchikov:1977nu}
\bibinfo{author}{\bibfnamefont{A.~B.} \bibnamefont{Zamolodchikov}}
  \bibnamefont{and} \bibinfo{author}{\bibfnamefont{A.~B.}
  \bibnamefont{Zamolodchikov}}, \bibinfo{journal}{Nucl. Phys.}
  \textbf{\bibinfo{volume}{B133}}, \bibinfo{pages}{525}
  (\bibinfo{year}{1978}{\natexlab{a}}).

\bibitem[{\citenamefont{Zamolodchikov and
  Zamolodchikov}(1978{\natexlab{b}})}]{Zamolodchikov:1977hh}
\bibinfo{author}{\bibfnamefont{A.~B.} \bibnamefont{Zamolodchikov}}
  \bibnamefont{and} \bibinfo{author}{\bibfnamefont{A.~B.}
  \bibnamefont{Zamolodchikov}}, \bibinfo{journal}{Phys. Lett.}
  \textbf{\bibinfo{volume}{B72}}, \bibinfo{pages}{481}
  (\bibinfo{year}{1978}{\natexlab{b}}).

\bibitem[{\citenamefont{Witten}(1978)}]{Witten:1977xv}
\bibinfo{author}{\bibfnamefont{E.}~\bibnamefont{Witten}},
  \bibinfo{journal}{Nucl. Phys.} \textbf{\bibinfo{volume}{B142}},
  \bibinfo{pages}{285} (\bibinfo{year}{1978}).

\bibitem[{\citenamefont{Morales}(2005)}]{Morales:2005}
\bibinfo{author}{\bibfnamefont{H.}~\bibnamefont{Morales}}, Ph.D. thesis,
  \bibinfo{school}{University of Kentucky} (\bibinfo{year}{2005}).

\bibitem[{\citenamefont{Klaiber}()}]{Klaiber:1968}
\bibinfo{author}{\bibfnamefont{B.}~\bibnamefont{Klaiber}},
  \emph{\bibinfo{title}{The {T}hirring model}}, \bibinfo{howpublished}{In {\it
  Lectures in Theoretical Physics}, edited by Asim O. Barut and Wesley E.
  Brittin (Gordon and Breach, New York, 1968), Vol. X-A, p. 141-176.}

\bibitem[{\citenamefont{Abdalla et~al.}(2001)\citenamefont{Abdalla, Abdalla,
  and Rothe}}]{Abdalla:1991vu}
\bibinfo{author}{\bibfnamefont{E.}~\bibnamefont{Abdalla}},
  \bibinfo{author}{\bibfnamefont{M.~C.~B.} \bibnamefont{Abdalla}},
  \bibnamefont{and} \bibinfo{author}{\bibfnamefont{K.~D.} \bibnamefont{Rothe}},
  \emph{\bibinfo{title}{Nonperturbative methods in two-dimensional quantum
  field theory}} (\bibinfo{publisher}{World Scientific},
  \bibinfo{address}{Singapore}, \bibinfo{year}{2001}), \bibinfo{edition}{2nd}
  ed., \bibinfo{note}{832 p}.

\bibitem[{\citenamefont{Schwinger}(1959)}]{Schwinger:1959xd}
\bibinfo{author}{\bibfnamefont{J.~S.} \bibnamefont{Schwinger}},
  \bibinfo{journal}{Phys. Rev. Lett.} \textbf{\bibinfo{volume}{3}},
  \bibinfo{pages}{296} (\bibinfo{year}{1959}).

\bibitem[{\citenamefont{Witten}(1984)}]{Witten:1983ar}
\bibinfo{author}{\bibfnamefont{E.}~\bibnamefont{Witten}},
  \bibinfo{journal}{Commun. Math. Phys.} \textbf{\bibinfo{volume}{92}},
  \bibinfo{pages}{455} (\bibinfo{year}{1984}).

\end{thebibliography}

\end{document}